**Godot is not coming: when we will let innovations enter psychiatry?**

Milena B. Čukić[1,2,3*]

[1] Institute for Technology of Knowledge, Complutense University, Madrid, Spain

[2] General Physiology with Biophysics Dpt., School of Biology, Belgrade University, Belgrade, Republic of Serbia

[3] 3EGA B.V., Amsterdam, the Netherlands

Autor Note

The Author of this manuscript is Milena B. Čukić, PhD, Koningin Wilhelminaplein 644, 1062KS Amsterdam, the Netherlands, phone:+31615178926 email: micukic@ucm.es

* Correspondence concerning this article should be addressed to Milena B. Čukić, PhD, Koningin Wilhelminaplein 644, 1062KS Amsterdam, the Netherlands, phone:+31615178926 email: micukic@ucm.es. Or micu@3ega.nl

**Conflict of interest statement:** the author declare no conflict of interest.

**Consent statement/Ethical approval**: Not required (this is cross section analysis based on published work)

**Author contributions:** Milena B. Čukić conceived the original idea and designed the overall structure of the manuscript; and wrote and reviewed the manuscript.

**Author ORCID:** Milena B. Čukić 0000-0002-9162-987X

**Acknowledgement**: This work is partially funded by RISEWISE (H2020-MSCA-RISE-2015-690874).

**Abstract**

Current diagnostic practice in psychiatry is not relying on objective biophysical evidence. Recent pandemic emphasized the need to address rising number of mood disorders (in particular, depression) cases in more efficient way. We are proposing several already developed practices that can help improve that diagnostic process: detection based on electrophysiological signals (both electroencephalogram and electrocardiogram based) that were shown to be accurate for clinical practice and several modalities of electromagnetic stimulation that were proven to ameliorate symptoms of depression. In this work we are connecting the two with explanation coming from physiological complexity studies (and our own work) as well as advanced statistical methods like machine learning and bayesian inference approach. It is shown that fractal and nonlinear measures can adequately quantify previously undetected changes in intrinsic dynamics of a physiological systems, providing the basis for early detection of depression. We are also advocating for early screening of cardiovascular risks in depression which is in connection to previously described decomplexification of autonomous nervous system resulting in symptoms recognised clinically. All that said, additional information about the level of complexity can help clinicians make better decision in therapeutic process, increase overall effectivity of the treatment, and finally increase the quality of life of the patient.

**Introduction**

The world is chaotic and unpredictable, and crisis are hitting when we do not expect them. This pandemic amplified and made visible some of the problems that did not attract that much attention before, like the existing crisis in psychiatry. That especially applies to mood disorders. Hence, we are going to focus on depression as the current leading cause of disability [1].

Pandemic revealed how fragile our mental healthcare systems are. Researchers are repeatedly reporting that due to lockdown, isolation, the fear of infection, economic hurdles, anxiety and prolonged lack of social contacts contributed to soaring number of people suffering from depression. The recent research showed that in December 2020 in USA more than 42% of people surveyed by the US Census Bureau in December reported symptoms of anxiety or depression. That is more than 11% increase when compared to the same month in 2019[2]. The same picture is similar worldwide. Before Covid-19 pandemics, WHO stated that approximately 350 million of people worldwide are treated for depression. Those who are offering online mental health support are reporting that the numbers are increasing even in the first months of COVID-19 crisis, bringing a "mental health tsunami" in many countries with an increasing rate higher than 48%[3].

Based on current practice, one can say that when it comes to diagnostics and treatment of depression, psychiatry does not have so much in common with modern medical scientific methods. The absence of objective (evidence-based) biomarkers and reliance solely on DSM/ICD as a tool for classification, and personal/biased impression/experience of the therapist, plus reliance on a self-report from the patient (that might be misleading or omitting important details) yield poor performance of overall treatment of mental disorders[4,5].

In their exhaustive book important for understanding today's psychiatry "Endocrine Psychiatry: Solving the Riddle of Melancholia", Shorter and Fink (2010) are leading us through the history of psychiatric treatments of the disease, mentioning several unsuccessful attempts of the introduction of physicochemical tests in diagnostic process [6]. Dr. Fink is known as a supporter of the comeback of electroconvulsive therapy as well as dexamethasone suppression test (DST) in endogenous depression (or melancholy) cases. By meticulously describing early studies and the history of the development of tests in the fifties from endocrine psychiatry point of view, he draws our attention to the similarities of different neurological disorders with cases of severe depression, improving our understanding of their roots. With this historical perspective we learn that "…hypercortisolemia was seen as a marker of the active phase of the melancholia syndrome". Shorter and Fink are pointing out that melancholia is treatable with electroconvulsive therapy (ECT) and/or tricyclic medication (prevention of suicide included), if differentiated from non-melancholic depressions (those without cortisol abnormality at their root). It is also showed that multi-site collaborative study STAR*D "… failed to identify an antidepressant benefit for the new selective serotonin reuptake inhibitor (SSRI) agents" [7]. Van der Kolk also demonstrated that in a long run, the effect of SSRI is slightly different from placebo-like intervention [8]. Based on STAR*D it was concluded that an average effectivity (success rate) of the antidepressant treatment is between 17% and 30%, which is quite modest result [9]. That reflects the fraction of those remitting within the first year of treatment. It is well-known that those who reach out for help are possibly half of those who do have depression symptoms; the rest are reluctant to report it due to the stigma attached to mental illness. It is also known that many people do not want to take medication for the rest of their lives: around 30% of patients never pick up prescribed medication after diagnosed as depressed [10]. It is not just 19th century diagnostics, where the medication is prescribed as in trial-and-error fashion; the process of recovery is very slow (relapses are frequent), expensive and unpredictable, and sometimes the psychotherapy can last for decades. For example, those who suffer from bipolar depressive disorder are in average, misdiagnosed (and treated) as unipolar depression for eight years [11], and prescribed antidepressants were repeatedly shown to actually deteriorate their condition [12]. To make things worse, there is no antidepressive drug without more or less serious side effects; some patients decide to stop taking medication in order to avoid having parkinsonism-like symptoms on top of their depression [13]. There is also an ongoing debate about the additional disruptive role of antidepressants [14,15] and previously unknown effects on sleep patterns and apnea [16]. De Kwaasteniet demonstrated that there is a problem with functional connectivity within fronto-lymbic system in MDD [17] and other researchers reported abnormal connectivity strength related to the severity of symptoms in depression [18]. There is also serious cardiovascular disease (CVD) risk present in depression, or should we say the *depression is* a serious CVD risk, which is currently unattended in clinical practice [19]. For example, in cardiology, once a person has myocardial infarction, he is immediately screened for depression, since their risk of mortality increase greatly if those two disorders combine (Dhar et al., 2016). There are a plenty of evidence that perturbed heart rate variability (HRV) is robust risk factor of CVD and overall increased mortality, but that finding is not taken care of in daily psychiatric practice [19-22].

The informative comparison from Shorter & Fink's book is that DST methodology has its limitations, but a positive test 'verifies the diagnosis of melancholia much as the EEG verifies the seizure disorder or the ECG a cardiac event'[6]. In one of the most inclusive review of the state of our understanding of causes and features of depression, regarding various avenues of research for biomarkers on biochemical, biological, structural, and behavioral level, Willner showed that some root misconceptions are still there causing fruitless outcomes of the current approach to therapy- since mono-amine hypothesis from fifties' not a single novel approach to treatment nor detection entered the field [23]. He also demonstrated in one of earlier publications that the brain treated with antidepressants is simply 'not in the same state as the healthy brain' [24]. Both Fink and Willner (and countless other authors in last three decades, including Stoyanov and Maes [25]) explained why the psychiatry is in such crisis today; the long lasting status-quo, guarding the field from entering any innovation (including DST several decades ago) made it so conservative that different avenues of research are very often in conflict and cannot be reconciled. The solution is probably coming from the technical sciences, but the fact that medical practitioners (those who are working with depression patients) are probably unaware of those advanced methods pose a problem here. Something that is known/demonstrated in statistical physics or artificial intelligence, or complex systems dynamics theory or information theory for decades is somehow still 'novel' in the eye of clinical psychiatrist. Physiological complexity, that use fractal and nonlinear dynamics to characterize the changes in physiological systems, is well based in literature, especially in numerous applications in cardiology [26]. Goldberger, Pincus, Peng, Hausdorf, Klonowski and others showed that deeply rooted classical (conventional) approach to analyzing electrophysiological signals might have misleading results; yet all professionals relying on any electrophysiology use for example, spectral measures because they are built in the software used for recording [26-30]. Physiological systems are complex, thus they do not obey linearity nor additive laws; even when we can describe them with relatively simple mathematical models, they typically generate unpredicted outcomes, since their sub-units interact and exhibit characteristic intrinsic dynamics that can be treated successfully with nonlinear analysis only [31]. The stereotypy of disease can be quantified as a loss of 'healthy' complexity which we can detect by analyzing any physiological signal; the illness is usually different in a way that the signal becomes much more predictable and exhibit oscillatory-like behavior. Hence the source of recognition. In depression (particularly in MDD) there are various neuroimaging findings that demonstrate how that system is different from the healthy one. Researchers showed that, regardless the cause of the depression (here we are interested in detection), there is a compensation-like behavior – a difference that can be detected probably due to a deep structural change (detected as aberrant functional connectivity) [17,32-37]. We do not aim at a grand theory-driven approach that can yield personalized model of a disease that includes all possible biomarkers, symptoms, self-report, and epidemiological data. We simply want to propose a data-driven solution that can be noninvasive, low-cost, and cloud-based to help clinicians improve their effectivity of overall treatment of depression, as an efficient answer to the crisis-caused huge increase in number of patients.

Now it is clear that nonlinear electrophysiology-based markers can play a similar role today, as Fink and colleagues proposed DST in last century. The detection and monitoring of mood disorders can be significantly improved using data-driven computational psychiatry approach [38,39,40, 41, 42, 43, 44, 45].

The goal of this opinion paper is reviewing the work based on various methods of detection of depression from electrophysiological data (electroencephalogram/EEG and electrocardiogram/ECG) that were proven to be noninvasive and characterized by high accuracy. Also, we will present the review of different well-based methods (excluding pharmacological and talk therapies) shown to be effective in the treatment of depression [46]. We also aim at connecting the nonlinear signal analysis approach with the following areas: detection/diagnostics, forecasting the patients' response to therapy, monitoring the outpatients – detecting the stages of the disease, and the application of non-pharmacological methods used in depression treatment (i.e., various types of electromagnetic stimulations).

**Neuroimaging findings related to depression**

In his 1948 'Cybernetics and Psychopathology', Norbert Wiener stated that … '(there is) …nothing surprising in considering the functional mental disorders as fundamentally diseases of the memory, of the circulating information kept by the brain in the active state, and of the long-time permeability of synapses' which is in line with recent research results in recurrent dynamics and functional integration of the brain. Wiener (1948) called depression, paranoia, and schizophrenia a "functional mental disorder" [47]. The recent findings of the physiological, structural, and functional mechanisms underlying depression confirmed the importance of disrupted functional connectivity within fronto-limbic system in depression [32-37].

Several studies reported aberrant connectivity in depression. In their fMRI study of medication-free patients with major depressive disorder (MDD), Grimm et al. (2007) showed the existence of hypoactivity in the left dorsolateral prefrontal cortex (DLPFC) and hyperactivity in the right DLPFC [48]. In a more recent fMRI study, Ge et al. (2019) confirmed that the decreased connectivity of the right intermediate hippocampus (RIH) with the limbic regions was a distinguishing feature for treatment-resistant depression [49]. On the other hand, several connectivity studies reported that functional connectivity exists between subgenual anterior cingulate cortex (ACC) and medial temporal lobe (MTL) in depression, as well as in hippocampus and amygdala [50]. Furman et al (2011) reported the aberrated frontostriatal functional connectivity in major depressive disorder (MDD) [51], and Horn (2010) reported the correlation between functional connectivity of pregenual anterior cingulate cortex (pgACC) and severity of anhedonia in MDD [52]. Bluhm examined the resting state default-mode network connectivity in early depression using a seed region of interest analysis [32] and confirmed decreased connectivity within the caudate nucleus. Their study showed significantly reduced correlation between precuneus/posterior cingulate cortex and the bilateral caudate in depression compared with controls. Berman examined connectivity of the default network specifically in the subgenual cingulate both on- and off-task, and the relationship between connectivity and rumination in MDD [34]. Their results showed characteristic higher functional neural connectivity between posterior cingulate cortex and subgenual cingulate cortex, but during rest periods only. Vederine et al. (2011) [33] and de Kwaasteniet et al. (2013) [17] elaborated on abnormal functional connectivity in the fronto-lymbic system. Using the combination of fMRI and functional anisotropy (FA), de Kwaasteniet confirmed that white matter integrity of the uncinate fasciculus was reduced, and that functional connectivity between the subgenual ACC and MTL was enhanced in MDD. De Kwaasteniet also identified the negative correlation between uncinate fasciculus integrity and subgenual ACC functional connectivity with the bilateral hippocampus in MDD but not in healthy controls; this negative structure-function relation was positively associated with depression severity [17].

Zhang and his colleagues (2011) published a fMRI/graph theory (small world) study confirming disrupted brain connectivity networks in drug-naïve first-episode MDD [35]. It seems that MDD disrupts the global topological organization of the whole-brain networks. There are studies that emphasized the disrupted brain connectivity in mental disorders [36, 53, 54].

Lee et al. (2011) tested the connectivity strength of resting state EEG as a potential biomarker of treatment response in major depressive disorder. They concluded that '…the stronger the connectivity strengths, the poorer the treatment response.' Their experiment also showed that frontotemporal connectivity strengths could be a potential biomarker to differentiate responders from slow responders and non-responders in MDD. Chen et al (2017) reported higher amplitude of low-frequency fluctuations (ALFF) in both the amygdala and hippocampus in participants with MDD compared to their healthy peers. Using graph theoretical analysis, they found that clustering coefficient, local efficiency, and transitivity are decreased in MDD patients [37]. In their reviews, Wayne et al. (2008) [56] and Willner et al. (2013) covered almost all aspects examined in the quest of understanding the characteristic features of depression [23]. The first one focused on structural and functional abnormalities and neurocircuitry in depression, and the second one more broadly reviewed the present and dominant approaches in this area of research. Prior to this review, Willner et al. (2005) also demonstrated that antidepressants do not normalize brain activity: 'mood and behavior are restored to normal, but antidepressant-treated brain is in a different state from the non-depressed brain' [24]. As a sum, all the changes found in depression indicate that the main characteristic of MDD is in their abnormal connectivity and transfer of information, rather than in solely physical differences. In their Granger Causality study about depression, Hamilton and colleagues are questioning the importance of the functional connections between candidate regions found to be abnormal in depression [55]. Their research relied on then handful number of prior studies that yielded information about cross-structural communication and influence in depression [57,58]. Based on previously confirmed aberrant interrelations in MDD, they applied multivariate Granger Causality to estimate the extent to which preceding neural activity in one or more seed regions predicted subsequent activity in target brain regions in the analysis of blood oxygen-level-dependent (BOLD) data.

Hamilton found that increased activity in ventral anterior cingulate cortex (vACC) could be predicted by the activation of hippocampus in patients with depression. In addition, the authors showed a mutual reinforcing effect between vACC and prefrontal cortex. Further, it was found that hypoactivity of dorsal cortical regions might be predicted with vACC and hippocampal increased activity. They demonstrated that aberrant patterns of effective connectivity implicate disturbances in the mesostriatal dopamine system in depression contributing to the knowledge about the primary role of limbic inhibition of dorsal cortex in the cortico-limbic relation [55]. It seems that many above mentioned areas probably illustrate different dynamics as networks, active in specific tasks known to be characteristically different in depression.

As an illustration on how the dynamics of healthy person's EEG is different from an MDD one, we present here one of our earlier unpublished findings (Figure 1). The dataset (a part of registered and approved study in 2015) used is described in several publications of our group [41-44]. We re-analyzed the dataset collected for another study [42] where we applied nonlinear analysis of EEG to differentiate between episode and remission of disease (21 MDD and 20 Healthy controls resting state EEG, all 19 electrodes 10/20 standard). Instead of nonlinear approach we analyzed the samples of the raw signal and applied Transfer entropy that can also yield information about the direction of information transfer (based on available free software MuTe)[59].

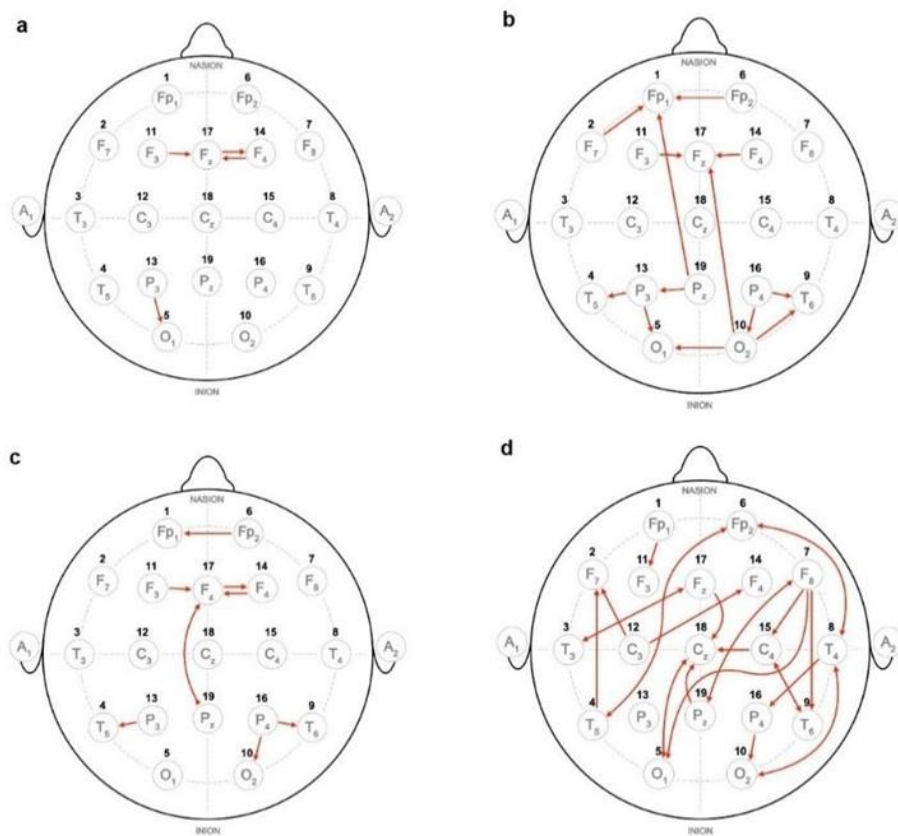

**Figure 1**. Results for BINNUE analysis. The figures a, b and c represent the analysis performed on segments 1, 2 and 3, respectively, taken from 5-minute EEG recording of patient group (those with MDD) with the standard 10/20 system. Figure d is a representation of BINNUE results for Healthy control group (HC). Standard names of electrode positions according to 10/20 system are Fp1 to Pz, but the bold number above (1-19) corresponds to the order of the channels analyzed in our analysis.

Our results show that the dynamics of healthy resting-state EEG is substantially different from the dynamics of MDD brain: the interactions (information transfers) in healthy controls are numerous during resting state, contrary to MDD brains which are repeatedly showing the "isolated" activity in frontal, parietal, and temporal areas. It should be emphasized that F3 (left DLPF)-Fz (frontal midline region)-F4 (right DLPFC) regions show engagement in each segment.

Knowing that right DLPFC region is involved in processing negative emotions (fear, anxiety, sadness), our results are in line with the previously described inability of persons with MDD to disengage from negative emotional content, as described in [23,34,60]. Graphical representation of transfer entropy (TE) in MDD shows just a few nodes (electrode positions) probably representing the abnormal cortical functional connectivity as a reflection of the one within the fronto-limbic system [33,17], engaged in negative information.

Recent research showed that TE is equivalent to Granger Causality (GC) for the data drawn from a Gaussian distribution [61,62]. TE can be seen as a difference of two conditional entropies [59] and can detect information transfer, discovering purely non-linear interactions between time-series under study. When we compare our findings with previously used GC results (on BOLD dataset see [55]), there are a certain number of connections that are in line with those findings. Hamilton et al. (2010) found that while observing moment to moment interactions, hippocampi were influencing vACC and consequently decreased activation of DLPFC [55]. It seems that hippocampus has a critical role in affecting depresotypic neural responses [49]. Of course, we cannot claim that we detected anything below the level of cortex by EEG, but the connections from Cz-Fp1 (Fig2/a), Pz-Fp1 (Fig1/b), and Pz-F2 (Fig1/c) are illustrating the direction of influences. It seems that F3, Fz and F4 are the most pronounced way of the information flow in our results. In this context we used EEG as the oldest form of a neuroimaging technique. To summarize, many other neuroimaging techniques showed that differences in MDD can be detected (MRI, fMRI, FA, DTI, PET) but due to very high cost of so sophisticated recordings all above mentioned studies usually rely on a modest sample. Computational psychiatry, a combination of computational neuroscience methods in psychiatry, started as a movement to improve the situation, sometimes around 2011/2012 (all the publications describing theory-driven approach started in 2013). What is more, we realized that, despite the very good results of Theory-driven computational psychiatry projects, they are simply not for everyone, since it is too expensive for frequent clinical checkups. From another point of view (machine learning), there is ongoing debate, whether the basic postulates of machine learning were satisfied in majority of neuroscientific studies relying for example, on fMRI or MRI scans, since the number of variables (features) is usually an order of magnitude larger than the number of participants, making the conclusions misleading [4,5,44,63] and generalization unreliable. Therefore, we conclude that further research would be potentially more useful if we focus on registration techniques that are present at any clinic (or even portable detections), easy and cheap to record, with minimal time consumption-namely EEG and/or ECG.

**Physiological complexity**

Electrophysiological signals are typically nonstationary, nonlinear, and noisy [30]. And are usually treated with some reductionist approach like analysis based on Fourier's analysis. As in many complex dynamics systems, physiological systems comprise of many sub-units that are coupled (i.e., they interact). Therefore, the signal that represents that complex dynamics (like EEG, ECG, EMG) do not obey linearity and superposition (additive principle). Hence, the application of any reductionist approach that assumes superposition and linearity will lead to erroneous conclusions and interpretations. For that reason, we suggest relying on fractal and nonlinear analysis application in electrophysiology.

It is well documented (especially in cardiology) that due to illness complex physiological signals exhibit a so-called paradoxical decrease of ordered variability, leading to the system becoming less able to appropriately adapt to fast changes of external (or internal) conditions. The consequence of this 'decomplexification' [64] is that a system as a whole becomes more predictable (less irregular) departing more and more from a healthy system's dynamics.

Dr. Reimann called the scientific attention to his observations about distinctive 'periodic diseases' (with characteristic cyclic patterns) in 1948, unsuccessfully. 50 years later,

Goldberger explained how that pathological complexity can be in line with stereotypy of disease [31]. Increased periodicity is associated with increased regularity (loss of complexity) and emergence of structured order and loss of variability (stereotypy). Pathological complexity, in case of MDD, actually means the loss of complex organized variability (characteristic for healthy dynamics) dictated by mechanisms of neural control that are proven to have fractal nature [26,28,30,65]. Many physiological functions like heart rate and human gait have a long-range correlation structure, while in mood disorders, a breakdown of scale-free dynamics can be observed. For the best review of applying such analysis of EEG with the aim of extracting depression features see de la Torre-Luque [66].

Physiological complexity, as scientific discipline, treats physiological phenomena with nonlinear dynamics methods stemming from statistical physics. It is already shown that complex dynamics systems are better conceptualized as part of dynamic reordering rather than as manifestation of disorderly processes [31]. Extensive research about fractal structure and function in physiology showed that scale-invariance is a central organizing principle. Moreover, fractal processes generate irregular fluctuations on multiple timescales that break down with aging and disease [67]. When a physiological system is perturbed, that break down of fractal organization can be quantified with nonlinear measures, like in heart dynamics, human gate, mood disorders, cardio-pulmonary processes, and other diseases. As if the affected person's system dynamics become less unique (stereotypy) constrained to single scale or frequency, losing background variability (decomplexification) present in healthy organism.

Once we quantify the pathological changes in electrophysiological signals (by fractal and nonlinear measures) describing their intrinsic dynamics, they can also be used as features for further machine learning models. As we already showed in our previous work [42,43,44] the proper nonlinear characterization of the signal leads to highly accurate classification of depression (and even the phases of the disease) with any of most popular machine learning models employed. When we approach machine learning with added

Bayesian principles (that proved to outperform standard frequencionistic thinking) the generalized model can revolutionize clinical diagnostics of early MDD.

**EEG based detection**

It became obvious that current 'gold standard' DSM/ICD taxonomies are unreliable constructs, and therefore unable to revise conventional classifications and help get valid predictions from clinical perspective [25].

Many researchers tried to identify how specific parameters present in EEG change with MDD; apart from some findings related to spectral analysis of standard EEG subbands (that did not produce a confirmation of their physiological significance), fractal and nonlinear analysis yielded important results [66] to help us understand how the changes of CNS due to illness are reflected in the electrical signal it produces. In line with above mentioned superior performance of nonlinear analysis (compared to conventional methods) it is reasonable to apply fractal and nonlinear methods of analysis when we study such a complex and nonlinear dynamical system as brain. Before researchers started using nonlinear analysis, EEG was mainly reserved for confirmation of epileptic foci in neurology; it was believed for a long time that the observed variability of a signal represents 'noise'. In fact, it is possible to extract a lots of valuable information from that 'noise', and pioneers of nonlinear dynamics application in cardiology even devised the method on how to identify whether an electrical signal has physiological origin (and meaning) or it is simply random noise [29, 64].

There are many families of fractal and nonlinear measures one can apply to study a physiological signal, but the crucial part is its interpretation that requires a specific knowledge of a context (what does regularity quantify?) [68]. Healthy system exhibits a complex nonlinear behavior, and it can swiftly/abruptly change its dynamics if the conditions (both external and internal) require that; it is as if in disease, the organism loose that ability to adapt to changes-which predictable behavior is consequence of.

Without entering discussion why, it is already confirmed that we can detect increased excitability on cortex of a person who is diagnosed with MDD from its EEG signal [39,40,42,43,44, 69]. Many researchers showed that whatever method is used, EEG of a patient diagnosed with depression appears more complex that the one recorded from a healthy control [44]. Let's say that our aim is at this point to detect whether a person has depression; this feature can be used for detection, since EEG is quite easy to record. Almost every modern hospital has the apparatus, and many novel portable devices are emerging on market. As de la Torre-Luque and Bornas concluded after reviewing many different results, difficult to compare in a methodological sense, it appears that the EEG signal recorded from depressed patient is more *random* than complex; as if it reflects the change in intrinsic dynamics of a system-which can be seen as decreased organized variability (if we use fractal analysis it is 'the roughness' of the signal) [66].

As an illustration we offer an example of the fact that there is a good separation of the data when the resting-state EEG is characterized with sample entropy (SampEn); on Figure 2 there is PCA (first three principal components of all calculated SampEn on 21 MDD/red patients' and HC's/blue EEG)

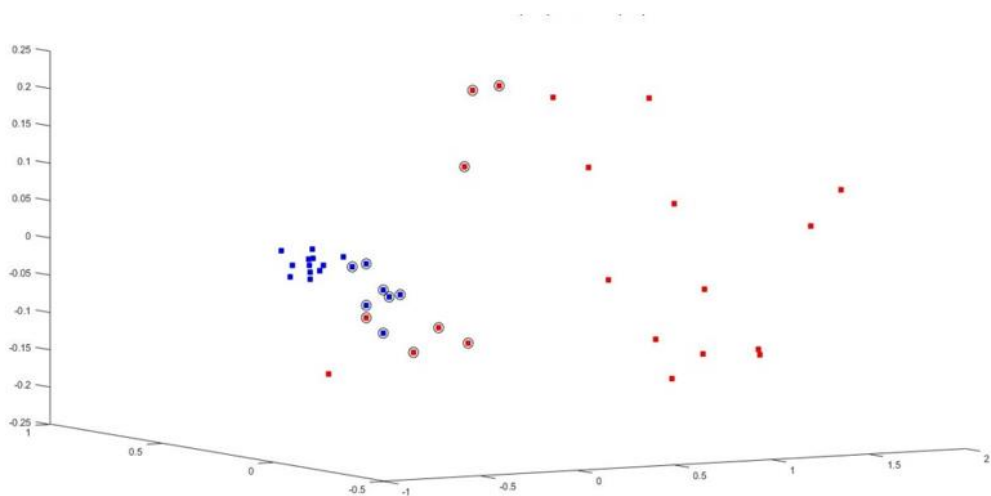

**Figure 2.** A good separability of the data illustrated with PCA applied on SampEn values calculated from resting-state EEG of MDD patients (blue) and Healthy controls (red). The healthy group is dispersed and exhibit greater variability compared to the MDD group (again, from work published in 41-44).

In our recent study [44] we confirmed the possibility of quantifying the difference between patients diagnosed with depression and healthy controls using HFD and SampEn on resting state EEG. For the first time it was shown that it is possible to differentiate patients who were in episode and remission phase.

There are several groups of research that were based on EEG: those that examined the changes when a person is presented with a stressful or a pleasant stimulae, those that compared the features of EEG before and after an intervention (like medication, some form of electromagnetic stimulation or a psychological therapy), those who tried to elucidate why the treatment was not successful (TRD) and those who used resting-state EEG in their analysis. As we believe that the analysis of the electrophysiological data should opt for minimal preprocessing (to prevent the loss of information) we chose to analyze artifact-free epochs of a resting-state EEG, with closed eyes; in our experience it yields the optimal results. Also, knowing that every mathematically different measure (and as we mentioned, there are vast families of nonlinear methods we can apply) could extract another information about the data, we also recommend calculation of several, mathematically different methods. In our case we combined Higuchi fractal dimension and some of entropy measures (like SampEn, Multiscale entropy and many other entropy-based measures), but other measures also showed to be successful, like Detrended fluctuation analysis (DFA), Recurrence plot analysis (RQA), Lempel-Ziv complexity, Largest Lyapunov exponent, Poincaré plots etc. This is the first step in analysis; it is just characterization of a signal (EEG) to extract the features that best describe it. The next step could be Feature selection-in order to decrease the dimensionality of a problem [70], like for example Principal Component Analysis (PCA) that we use in our machine learning projects.

We showed that once you characterize the signal in appropriate way, any of most popular machine learning models yield reasonably high accuracy in depression detection classification task [43]. That is the confirmation of our first presumption that the nonlinear, nonstationary, and noisy signals (and all physiological signals are) is best described by fractal and nonlinear analysis. When one combines spectral analysis to extract the features, the result is much less successful classification and the generalization of that model might be misleading. We dealt with methodological issues in more detail in another publication [44].

There are also research projects that combined EEG features and prior medical data (and other epidemiological data of MDD patients) to improve detection, or they combined other imaging (MRI), medical data and early childhood adverse events [71], or they performed machine learning to navigate better therapy even without those features [72]. There are also various applications of deep learning for depression detection task and various applications of unsupervised learning, but our point here is that a good approach to treating this detection is to first characterize the signal with fractal and nonlinear measures and then combine it with some form of advanced statistical methods, from data mining, machine learning to deep learning and even the most novel Bayesian approaches to machine learning (sometimes called Bayesian or Active inference). To summarize, we believe that relying solely on EEG for detection could be one of the potentially useful methods, since it also complies with GDPR legislature, hence no personal data are needed (the identity cannot be reconstructed from electrophysiological data that is).

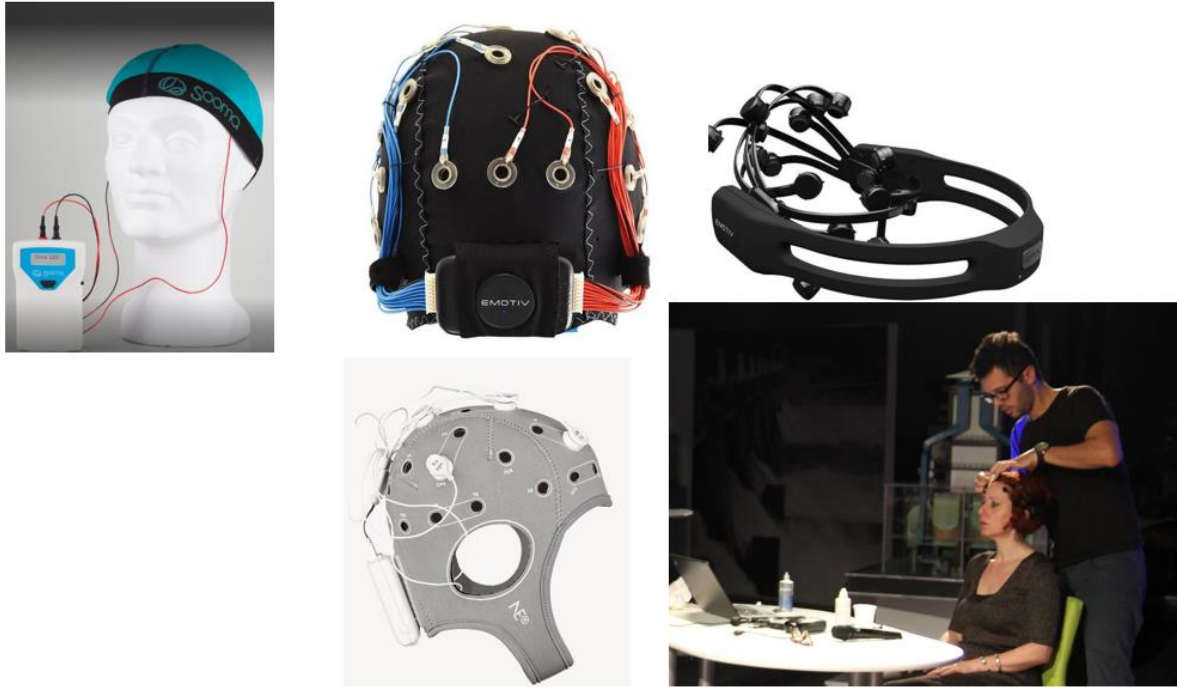

**Figure 3.** Illustration of possible EEG recordings, standard and portable (on this picture there are two members of our team, the author of this paper included, not the patients. The photo was made during our public demonstration during SUTRA festival of science in Republic Serbia in 2019, https://www.youtube.com/watch?v=pHAFflliOTo&ab_channel=Milena%C4%8Cuki%C4%87).

In addition to described methods of detection, there are portable EEG caps that are at the moment mainly reserved for research and are quite expensive, but we expect them to become more available to general population in near future and make this detection more applicable and accessible (for example for monitoring at home solution).

**ECG-based detection**

The strong characteristic of various physiological signals in illness is that they are less complex then the healthy one, and they actually loose the background organized variability of a signal; if you compare for example ECG of a healthy person and someone who have congestive heart failure the signal is varying around almost the same mean value, but the second one exhibit such an oscillatory behavior that it is highly predictable, contrary to the healthy one. One of the best examples of detection from early physiological complexity research is clear distinction based on ECG to detect sudden infant death syndrome in newborns (SIDS) [73,74].

We now understand the connection between the dynamics of heart rate and autonomous nervous system (ANS), and we also know that it is characteristically aberrated in depression [75]. Nevertheless, in clinical psychiatry the risk of cardiovascular diseases (CVD) in depression [19,20] is still not attended. Variability heart rate is a robust predictor of CVD and overall mortality in depression [21], but patients are not yet screened for this. As with EEG based detection [43] it is possible to detect depression based on ECG signals recorded from MDD patients [76,77] but this method is not yet translated to practice. Again, the combination between nonlinear measures that are good at quantifying the electrophysiological signals, proved to be excellent features for later highly accurate machine learning classification [78,79]. Byun and colleagues found that entropy features are lower in the MDD group, compared to controls [78,79]. The measures used were approximate entropy (AppEn), Sample entropy (SampEn), Fuzzy entropy (FuzzEn) and Shannon entropy (ShanEn). They used leave-one-out procedure for feature selection and within it several classifiers (SVM, KNN, LDA and NB). They experimented with total 20 different entropy features and mentioned classifiers to complete depression detection task; similar to our work [43] all combinations yielded decent accuracies in detection (74,4%). For this study the sample comprised of 33 MDD patients and 33 age matched healthy controls: the signal they used were recorded from the wrists, not from the chest, but turned out to be sufficient for detection. The authors concluded that monitoring patients especially in the state of arousal would be ideal for detection. But from our own work we learned that the resting state electrophysiological signals are most information rich, which is in line with conclusions of Goldberger group and others [64]. In another study of the same group [78], authors detected major depressive disorder from linear and nonlinear heart rate variability features during mental task protocol.

**Monitoring of outpatients: detecting the phases/stages of the disease, forecasting the output of the therapy**

We recently made the large cross section analysis of this avenue of research (in preparation) and discovered that in last thirty years there are plenty of evidence that not only the mortality risk in depression can be easily assessed, but it is also possible to differentiate between melancholic and non-melancholic depression, which is utterly important for management of therapy [21]. Previously we demonstrated that it is possible (based on EEG) to differentiate between the phases of the disease, episode, and remission [42], but it turned out it is promising further direction for staging the disease from ECG, too (our yet unpublished data). The core of both methods is to use nonlinear measures for characterizing the electrophysiological signal, because it proved to be highly accurate feature extraction [44].

There are large body of evidence how this low-cost detection can be utilized in clinical practice simply as a reuse of already existing equipment (like standard EEG and ECG devices) or additional application of many portable monitoring devices that are already used in many Telehealth & IoT applications, that during this crisis demonstrated how helpful they are especially when a surge of novel patients are happening [3].

Depression is probably rare example when people are drawn to suicide, but many clinical attempts fail to forecast that fatal danger. Again, it is demonstrated that due to that aberration of ANS in depression it is possible to register the warning signs and prompt those who are taking care of patient, like in [80] where ECG based detection is used to confirm suicide ideation in MDD. There are also other proofs that only relying on increased heart rate and decreased heart rate variability- HRV [19] researchers confirmed the risks of suicide in analyzing aggregated medical data at large clinic in Paris can be extracted from ECG, BMI, and waist circumference, from patients who were screened for general health risks [81]. Byuin showed that we do not even have to use standard equipment, wristbands (or other portable devices) to perform the task [76, 78,79, 82].

Saad and colleagues published a quality study (2019) where they performed detection of specific heart-rate patterns extracted from ECG recorded during the sleep, with an accuracy of 79.9%. The algorithm was trained on a sample of 644 depression patients (and 529 controls) and tested on a group of 174 persons, comprising of both patients and healthy controls. Since the details of algorithm are protected as patent (by MediBio) no details about the methodology were disclosed in this publication. Thirteen years after Pincus published his conclusion that the most information rich electrophysiological signal for depression detection would be ECG while sleeping (since majority of those suffering from depression experience sleep disturbance) find its realization in this group project (beside Migliorini's group in 2012). What we concluded though is that Saad and colleagues did not use nonlinear measures, but spectral ones. It is important to mention that this research group recorded somnograms and collected ECG signal via portable devices [77].

It is also possible to forecast whether a person is a responder to certain therapy, for example a certain class of drug, or for ECT, or rTMS in cases of severe depression [83,84].

To summarize, it is possible to detect CVD risks in depression (which is the most missing detection at the moment), it is possible to confirm the depression before the prescription of antidepressants (that all have some side-effects[16]), it is possible to detect whether you should send a person to some form of electromagnetic stimulation (like repetitive transcranial magnetic stimulation or electroconvulsive therapy), it is possible to differentiate episode from remission (to safely change the medication), it is possible to confirm suicide-ideation and that all leads us to the question begging the answer: *why clinicians are reluctant to use it*? All those findings can save the psychiatry from obvious unsustainability and help clinicians to treat their patients in much safer evidence-based way.

**Electromagnetic stimulation as therapy**

Given the robust evidence in present body of literature, we can say that one of the characteristic of depression is that corresponding electrophysiological (both EEG and ECG) data reflect certain aberration of intrinsic dynamics, which can be used for detection and monitoring. That can be a decision support method for clinicians to navigate better future therapy and it would be the cheapest form of personalized approach to depression treatment.

It is also demonstrated in numerous studies that there are several methods of electromagnetic stimulation proven to help even in cases of treatment-resistant depressions ('resistant' relates to medication and psychotherapy or psychological interventions). Among those are electroconvulsive therapy (ECT), transcranial direct current stimulation (tDCS), repetitive transcranial magnetic stimulation (rTMS), vagus nerve stimulation (VNS). Although it is not strictly electrical stimulation, we might add a HRV biofeedback therapy to this list as a therapy proven to minimize CVD risk in depression.

Intervention studies from previous period are sparse, and among first attempts to explore this are two studies which examined the effect of Electroconvulsive Therapy (ECT) on VHR in serious MDD. Shultz et al., (1997) [85] and Bozkurt et al., (2013) [83] published their results on improvement of decreased variability heart rate in very serious depression's BDI after ECT, but they used classical spectra methodology for analysis. Gandahar and colleagues (1999) [86], contrary to them used fractal dimension of EEG to demonstrate the impact of ECT, although without analyzing the heart rhythm. Recently, another two research groups applied similar approach to potentially forecast the responder patients to electromagnetic stimulation [84,87]. They both find a straight-forward connection between positive outcome of the rTMS therapy and evident changes in complexity in EEG signals of MDD patients (comparison was made between responders and non-responders on therapy). As in our recent study [44], they demonstrated the connection between complexity changes and effectivity of therapeutic role of rTMS/tDCS in MDD.

What is important to stress here is the fact that various electromagnetic stimulation techniques help alleviate the symptoms in TRD, is connected to the insights we have from Physiological complexity results from depression research. It is confirmed that depression is characterized by increased complexity of EEG (increased excitability possibly due to deep white matte changes like in [17] by numerous methodologically different methods [66]. Another important information is that antiepileptic drugs are frequently helpful in MDD, hence, there is a need and a therapeutic goal to decrease increased physiological complexity. All electromagnetic stimulation treatments are doing precisely that: decrease the present increased complexity, which is a 'hallmark' of, depression [44]. The effectivity of ECT is the highest, but patients are reluctant to accept it possible due to strongly negative media image of this treatment. Recent reviews [88,89] confirmed that both rTMS (37%) and tDCS (42%) are effective in higher percentage than any of present classes of antidepressants (~max 26-27%) regularly prescribed for depression treatment. There are also additional modalities of those two methods of stimulation, but their effectivity is not yet known in detail, due to recent application start [88]. The limitation of both tDCS and rTMS (as well as ECT before them) that their positive effect on patient's level of symptoms is temporary; they all need maintenance. It is interesting though, that tDCS that now can be applied in ambulance or even at home on a daily basis (portable tDCS, for example Sooma Medical) they can serve as a maintenance method for much stronger rTMS. The period of efficiency is usually individual but can be (depending on source) around six to eight months; then it needs to be reapplied. Both rTMS and tDCS are noninvasive, although we think a new interpretation of that term is sorely needed. When the term was coined, 'non-invasive' meant that the stimulation electrode is not entering into the crania; but based on other findings since then we can hardly say they are completely non-invasive in contemporary sense of the word[46]. We found evidence that even a very weak direct current application (as in tDCS) can leave influence more than half an hour after the application, looking at one of our PCA analyses of raw signal (EEG) samples. Figure 4 depicts how separable are the time trajectories in phase-space before and 30 minutes after the tDCS stimulation (one of previously not reported illustrations of a research published in [45]).

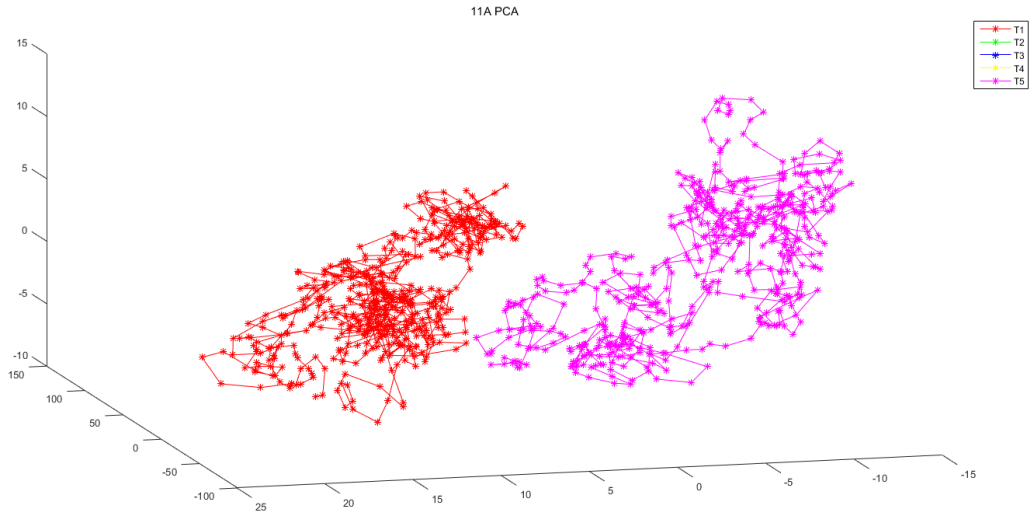

**Figure 4.** PCA shows the separation of trajectories in phase-space reconstructed from raw signal samples (EEG) before (red) and half an hour after (purple) the tDCS stimulation (previous publication that used recycled/already published data) [44].

**Conclusion**

Beside existence of various methods developed within technical sciences, there is a reluctance to their application in everyday psychiatry. After Covid-19 crisis demonstrated our mental health care systems so fragile, the time is ripe for application of demonstrated methods to introduce objective biomarkers in clinical diagnostics. To paraphrase Keyns's words – 'acceptance of new ideas is slow not because the new ideas are absent, but because of our inability to leave behind old ones we used for such a long time'.

**References**

1. **WHO** 2017 https://www.who.int/news-room/fact-sheets/detail/depression

2. **Abbott A**. COVID's mental-health toll: how scientists are tracking a surge in depression. *Nature* 2021; **590**: 194-195 (2021). doi: 10.1038/d41586-021-00175-z.

3. **Inkster B**, Digital Mental Health Data Insights Group (DMHDIG). Early Warning Signs of a Mental Health Tsunami: A Coordinated Response to Gather Initial Data Insights From Multiple Digital Services Providers. *Front Digit Health* 2021; **2**: 578902. doi: 10.3389/fdgth.2020.578902

4. **Whelan R**, Garavan H. When optimism hurts: inflated predictions in psychiatric neuroimaging. *Biol Psychiatry* 2014; **75**: 746-748. doi: 10.1016/j.biopsych.2013.05.014.

5. **Gillan CM**, Whelan R. What big data can do for treatment in psychiatry. *Current Op Behav Sci* 2017; **18**: 34-42. doi: 10.1016/j.cobeha.2017.07.003.

6. **Shorter E**, Fink M. *Endocrine psychiatry: Solving the riddle of melancholia*. New York: Oxford University Press, 2010.

7. **Gillan CM**, Daw ND. Taking psychiatry research online. *Neuron (NeuroView)* 2016; **91**: 19–23.

8. **van der Kolk BA**, Spinazzola J, Blaustein ME, Hopper JW, Hopper EK, Korn DL, Simpson WB. A randomized clinical trial of eye movement desensitization and reprocessing (EMDR), fluoxetine, and pill placebo in the treatment of posttraumatic stress disorder: treatment effects and long-term maintenance. *J Clin Psychiatry* 2007; **68**: 37-46. PMID: 17284128.doi: 10.4088/jcp.v68n0105. PMID: 17284128.

9. **Rush AJ**, Trivedi MH, Wisniewski SR, Nierenberg AA, Stewart JW, Warden D, Niederehe G, Thase ME, Lavori PW, Lebowitz BD, McGrath PJ, Rosenbaum JF, Sackeim HA, Kupfer DJ, Luther J, Fava M. Acute and longer-term outcomes in depressed outpatients requiring one or several treatment steps: a STAR*D report. *Am J Psychiatry* 2006; **163**:1905-17. doi: 10.1176/ajp.2006.163.11.1905. PMID: 17074942.

10. **Williams M**. Penman D. The new psychology of Depression. *Open Culture*. University of Oxford Podcasts. 2016http://podcasts.ox.ac.uk/series/new-psychology-depression

11. **Singh T**, Rajput M. Missdiagnosis of Bipolar Dysorder. *Psychiatry (Edgmont)* 2006; **3**: 57-63.

12. **Patel R**., et al. Do antidepressants increase the risk of mania and bipolar disorder in people with depression? A retrospective electronic case register cohort study. *BMJ Open* 5(12) 2015. DOI: 10.1136/bmjopen-2015-008341.

13. **Sacks O**. Hallucinations. *Picador*. 2012.

14. **Licht CM**, de Geus EJ, van Dyck R, Penninx BW. Association between anxiety disorders and heart rate variability in the Netherlands Study of Depression and Anxiety (NESDA). *Psychosom Med* 2009; **71**: 508-518.

15. **Licht CM**, Penninx BW, de Geus EJC. To include or not to include? A response to the meta-analysis of heart rate variability and depression. *Biol Psychiatry* 2011; **69**: e1; author reply e3–4. doi:10.1016/j.biopsych.2010.06.034.

16. **Robillard R**, Dion K, Pennestri MH, Solomonova E, Lee E, Saad M, Murkar A, Godbout R, Edwards JD, Quilty L, Daros AR, Bhatla R, Kendzerska T. Profiles of sleep changes during the COVID-19 pandemic: Demographic, behavioural and psychological factors. *J Sleep Res* 2021; **30**: e13231. doi: 10.1111/jsr.13231. Epub 2020 Nov 17. PMID: 33200477.

17. **de Kwaasteniet B**, Ruhe E, Caan M, Rive M, Olabarriaga S, Groefsema M, Heesink L, van Wingen G, Denys D. Relation Between Structural and Functional Connectivity in Major Depressive Disorder. *Biol Psychiatry* 2013; **74**:40-47. DOI: 10.1016/j.biopsych.2012.12.024.

18. **Lee TW**, Wu YT, Yu YWY, Chen MC, Chen TJ. The implication of functional connectivity strength in predicting treatment response of major depressive disorder: A resting EEG study. *Psychiat Res Neuroimaging* 2011; **194**: 372-377. DOI: 10.1016/j.pscychresns.2011.02.009.

19. **Kemp AH**, Kemp DS, Quintana MA, Gray KL, Felmingham KB, Gatt JM. Impact of Depression and Antidepressant Treatment on Heart Rate Variability: A Review and Meta-Analysis. *Biol Psychiatry* 2010; **67**: 1067-1074. doi:10.1016/j.biopsych.2009.12.012.

20. **Kemp A**. Depression, antidepressant treatment and the cardiovascular system. *Acta Neuropsychiatrica* 2011; **23**: 82-83. doi:DOI 10.1111/j.1601-5215.2011.00535.x.

21. **Kemp AH**, Quintana DS, Quinn DR, Hopkinson P, Harris AWF. Major depressive disorder with melancholia displays robust alterations in resting state heart rate and its variability: implications for future morbidity and mortality. *Front Psychol* 2014; doi: 10.3389/fpsyg.2014.01387

22. **Koch C**, Wilhelm M, Salzmann S, Rief W, Euteneuer F. A meta-analysis of heart rate variability in major depression. *Psychol Med* 2019; DOI: 10.1017/S0033291719001351.

23. **Willner P**, Scheel-Krüger J, Belzung C. The neurobiology of depression and antidepressant action. *Neurosci Biobehav Rev* 2013; **37**: 2331-2371. DOI: 10.1016/j.neubiorev.2012.12.007.

24. **Willner P**, Hale AS, Argyropoulos SV. Dopaminergic mechanism of antidepressant action in depressed patients. *J Affect Disord* 2005; **86**: 37-45. DOI: 10.1016/j.jad.2004.12.010.

25. **Stoyanov D**, Maes MH. How to construct neuroscience-informed psychiatric classification? Towards nomothetic networks psychiatry. *World J Psychiatry* 2021; **11**: 1-12. doi:10.5498/wjp.v11.i1.1.

26. **Goldberger AL**, Peng CK, Lipsitz LA. What is physiologic complexity and how does it change with aging and disease? *Neurobiol Aging* 2002; **23**: 23-26.

27. **Pincus SM**. Approximate entropy as a measure of irregularity for psychiatric serial metrics. *Bipolar Disorders* 2006; **8**: 430–440. http://dx.doi.org/10.1111/j.1399-5618.2006.00375.x.

28. **Peng CK**, Hausdorff J, Goldberger A. Fractal mechanisms in neural control: Human heartbeat and gait dynamics in health and disease. In: Walleczek J, editor. *Self-Organized Biological Dynamics and Nonlinear Control*. Cambridge: Cambridge University Press, 1999.

29. **Hausdorf JM**., Peng CK., Ladin Z. Is walking a random walk? Evidence for long-range correlations in the stride interval of human gait. *J. Appl. Physiol*. 78:349-358, 1995.

30. **Klonowski W**. From conformons to human brains: an informal overview of nonlinear dynamics and its applications in biomedicine. *Nonlinear Biomed. Phys.* 2007; **1**: 5. doi: 10.1186/1753-4631-1-5.

31. **Goldberger AL**. Fractal Variability Versus Pathologic Periodicity: Complexity Loss and Stereotypy in Disease. *Perspect Biol Med 1997*; **40**: 543-561.

32. **Bluhm R**, Williamson P, Lanius R, Theberge J, Densmore M, Bartha R, Neufeld R, Osuch E. Resting state default-mode network connectivity in early depression using a seed region-of-interest analysis: Decreased connectivity with caudate nucleus. *Psychiat Clin Neurosci* 2009; **63**: 754-761. DOI: 10.1111/j.14401819.2009.02030.x.

33. **Vederine FE**, Wessa M, Leboyer M, Houenou JA. Meta-analysis of whole-brain diffusion tensor imaging studies in bipolar disorder. *Prog Neuropsychopharmacol Biol Psychiatry* 2011; **35**:1820-1826. DOI: 10.1016/j.pnpbp.2011.05.009.

34. **Berman MG**, Peltier S, Nee DE, Kross E, Deldin PJ, Jonides J. Depression, rumination and the default network. *Soc Cogn Affect Neurosci* 2011; **6**: 548-555. doi: 10.1093/scan/nsq080.

35. **Zhang J**, Wang J, Wu Q, Kuang W, Huang X, He Y, Gong Q. Disrupted Brain Connectivity Networks in Drug-Naive, First-Episode Major Depressive Disorder. *Biol Psychiatry* 2011; **70**: 334-342. DOI: 10.1016/J.BIOPSYCH.2011.05.018;

36. **Kim D**, Bolbecker AR, Howell J, Rass O, Sporns O, Hetrick WP, Breier A, O'Donnell BF. Disturbed resting state EEG synchronization in bipolar disorder: a graph-theoretic analysis. NeuroImage Clin 2013; **2**: 414-423. DOI: 10.1016/j.nicl.2013.03.007.

37. **Chen X**, Yang R, Kuang D, Zhang L, Lv R, Huang X, *et al.* (2017): Heart rate variability in patients with major depression disorder during a clinical autonomic test. *Psychiatry Res* 2017; **256**: 207-211.

38. **Ahmadlou M**, Adeli H, Adeli A. Fractality analysis of frontal brain in major depressive disorder. *Int J Psychophysiol* 2012; **85**: 206-211. DOI: 10.1016/j.ijpsycho.2012.05.001.

39. **Hosseinifard B**, Moradi MH, Rostami R. Classifying depression patients and normal subjects using machine learning techniques and nonlinear features from EEG signal. *Comput Methods Programs Biomed* 2014; **109**: 339–345

40. **Bachmann M**, Päeske L, Kalev K, Aarma K, Lehtmets A, Ööpik P, Lass J, Hinrikus H. Methods for classifying depression in single channel EEG using linear and nonlinear signal analysis. *Comput Methods Programs Biomed* 2018; **155**:11-17.

41. **Čukić M**, Pokrajac D, Stokić M, Simić S, Radivojević V, Ljubisavljević M. EEG machine learning with Higuchi's fractal dimension and Sample Entropy as features for successful detection of depression. arXiv 2018:-

42. **Čukić M**, Stokić M, Radenković S, Ljubisavljević M, Simić S, Savić D. Nonlinear analysis of EEG complexity in episode and remission phase of recurrent depression. *Int J Res Meth Psychiatry* 2019; e1816. DOI: 10.1002/MPR.1816

43. **Čukić M**, Stokić M, Simić S, Pokrajac D. The successful discrimination of depression from EEG could be attributed to proper feature extraction and not to a particular classification method. *Cog Neurodyn* 2020; DOI: 10.1007/s11571-020-09581-x.

44. **Čukić Radenković M**, Lopez V, Pavon J. Machine Learning Approaches for Detecting the Depression from Resting-State Electroencephalogram (EEG): A Review. *J Med Internet Res* 2020; **22**: e19548. DOI:10.2196/19548.

45. **Čukić M**, Stokić M, Radenković S, Ljubisavljević M, Pokrajac D. The Shift in brain-state induced by tDCS: an EEG study. In: Novel Approaches in Treating Major Depressive Disorder, ISBN: 978-1-53614-382-9, NOVA Scientific Publishers ltd, New York, USA, 2019.

46. **Čukić M**. The Reason Why rTMS and tDCS Are Efficient in Treatments of Depression. *Front Psychol* 2020; **10**: 2923. DOI: 10.3389/fpsyg.2019.02923.

47. **Wiener N**. *Cybernetics: Or Control and Communication in the Animal and the Machine*. Paris, (Hermann & Cie) & Camb. Mass. (MIT Press) ISBN 978-0-262-730099; 2nd revised ed, 1948.

48. **Grimm S**, Schmidt CF, Bermpohl F, Heinzel A, Dahlem Y, Wyss M, Hell D, Boesiger P, Boeker H, Northoff G. Segregated neural representation of distinc emotion dimensions in the prefrontal cortexand fMRI study. *Neuroimage* 2006; **30**:325-340.

49. **Ge R**, Torres I, Brown JJ, Gregory E, McLellan E, Downar JH, Blumberger DM, Daskalakis ZJ, Lam RW, Vila-Rodriguez F. Functional disconnectivity of the hippocampal network and neural correlates of memory impairment in treatment-resistant depression. *J Affect Disord* 2019; **253**: 248-256. DOI: 10.1016/j.jad.2019.04.096.

50. **Pezawas L**, Meyer-Lindenberg A, Drabant EM, Verchinski BA, Munoz KE, Kolachana BS, Egan MF, Mattay VS, Hariri AR, Weinberger DR. 5-HTTLPR polymorphism impacts human cingulate-amygdala interactions: a genetic susceptibility mechanism for depression. *Nat Neurosci* 2005; **8**: 828-834.

51. **Furman DJ**, Hamilton JP, Gotlib IH. Frontostriatal functional connectivity in major depressive disorder. *Biol Mood Anxiety Disord* 2011; **1**:11. DOI: 10.1186/2045-5380-1-11.

52. **Horn DI**, Yu C, Steiner J et al. Glutamatergic and resting-state functional connectivity correlates of severity in major depression – the role of pregenual anterior cingulate cortex and anterior insula. *Front Syst Neurosci* 2010; **4**: 33. DOI: 10.3389/fnsys.2010.00033.

53. **Van Essen DC**, Ugurbil K, Auerbach E et al. The Human Connectome Project: A data acquisition perspective. *Neuroimage* 2012; **62**: 2222-2231. DOI: 10.1016/j.neuroimage.2012.02.018.

54. **Castellanos FX**, Di Martino A, Craddock RC, Mehta AD, Milham MP. Clinical applications of the functional connectome. Neuroimage 2013; **80**: 527-540.

55. **Hamilton JP**, Chen G, Thomason ME, Schwartz ME, Gotlib IH. Investigating neural primacy in Major Depressive Disorder: multivariate Granger causality analysis of resting-state fMRI time-series data. *Mol Psychiatry* 2011; **16**: 763-772.

56. **Wayne C**, Drevets JLP, Furey M. Brain structural and functional abnormalities in mood disorders: implications for neurocircuitry models of depression. *Brain Struct Funct* 2008; **213**: 93-118. DOI: 10.1007/s00429-008-0189-x.

57. **Lozano AM**, Mayberg HS, Giacobbe P, Hamani C, Craddock RC, Kennedy H. Subcallosal cingulate gyrus deep brain stimulation for treatment-resistant depression. *Biol Psychiatry* 2008; **64**: 461-467.

58. **Seminowicz DA**, Mayberg HS, McIntosh AR, Goldapple K, Kennedy S, Segal Z et al. Limbic-frontal circuitry in major depression: a path modeling metanalysis. Neuroimage 2004; **22**:409-418.

59. **Montalto A**, Faes L, Marinazzo D. MuTe: A Matlab toolbox to compare established and novel estimators of the multivariate transfer entropy. *PloS one* 2014; **9**(10): e109462. DOI: 10.1371/journal.pone.0109462.

60. **Gotlieb IH**, Joorman J. Cognition and Depression: Current Status and Future Directions. Annu Rev Clin Psychol 2010; **6**:285-312. DOI: 10.1146/annurev.clinpsy.121208.131305.

61. **Barnett L**, Barrett AB, Seth AK. Granger causality and transfer entropy are equivalent for gaussian variables. Phys Rev Lett 2009; **103**: 238701.

62. **Hlavačkova-Schindler K**. Equivalence of granger causality and transfer entropy: A generalization. *App Math Sci* 2011; **5**: 3637–3648.

63. **Yahata N**, Kasai K, Kawato M. Computational neuroscience approach to biomarkers and treatments for mental disorders. *Psychiatry Clin Neurosci* 2017; **71**: 215-237 [FREE Full text] [doi: 10.1111/pcn.12502].

64. **Goldberger A**, Amaral LAN, Glass L, Hausdorff J, Ivanov PC, Mark RG, *et al.* 'PhysioBank, PhysioToolkit, and Phys-ioNet: Components of a new research resource for complex physiologic signals'. *Circulation* 2000; **101**: e215–e220. http://circ.ahajournals.org/cgi/content/full/101/23/e215.

65. **Eke A**, Herman P, Kocsis L, Kozak LR. Fractal characterization of complexity in temporal physiological signals. *Physiol Meas* 2002; **23**: R1-R38.

66. **De la Torre-Luque A**, Bornas X (2017): Complexity and Irregularity in the Brain Oscillations of depressive Patients: A Systematic Review. *Neuropsychiatry (London)* **5**: 466-477.

67. **Goldberger A** et al. (2002): Fractal dynamics in physiology: Alterations with disease and aging PNAS 99(1): 2466–2472.

68. **Pincus SM**, Goldberger AL. Physiological time-series analysis: what does regularity quantify? *Am J Physiol* 1994; **266**(4 Pt 2): H1643-1656.

69. **Ahmadlou M**, Adeli H, Adeli A. Spatiotemporal analysis of relative convergence of EEGs reveals differences between brain dynamics of depressive women and men. *Clin EEG Neurosci* 2013; **44**: 175-181. doi: 10.1177/1550059413480504.

70. **Witten IH**, Frank E, Hall MA. Measuring the social interactions of people with traumatic brain injury and their communication partners: The adapted Kagan scales. In: *DATA MINING Practical Machine Learning Tools and Techniques*. New York, USA: Elsevier; 2005.

71. **Tokuda T**, Yoshimoto J, Shimizu Y, Okada G, Takamura M, Okamoto Y, et al. Identification of depression subtypes and relevant brain regions using a data-driven approach. *Sci Rep* 2018; **8**: 14082. doi: 10.1038/s41598-018-32521-z.

72. **Checkroud A. M**. et al., Cross-trial prediction of treatment outcome in depression: a machine learning approach, *Lancet Psychiatry* 3(3) (2016): 243-250.

73. **Pincus SM** *et al.* Differentiation of women with premenstrual dysphoric disorder, recurrent brief depression, and healthy controls by daily mood rating dynamics. *J Psychiatr Res* 2008; **42**: 337-347.

74. **Pincus SM**, Cummins TR, Haddad GG. Heart rate control in normal and aborted SIDS infants. *Am J Physiol Regulatory Integrative Comp Physiol* 1993; **264**: 638-646.

75. **Rottenberg J**. Cardiac vagal control in depression: a critical analysis. *Biol Psychiatry* 2007; 74: 200-211. doi: 10.1016/j.biopsycho.2005.08.010.

76. **Migliorini M**, Mendez MO, Bianchi AM. Study of heartrate variability in bipolar disorder: linear and non-linear parameters during sleep. *Front Neuroeng*, 2012. doi: 10.3389/fneng.2011.00022.

77. **Saad M**. Using heart rate profiles during sleep as a biomarker of depression. *BMC Psychiatry* 2019; **19**: 168. https://doi.org/10.1186/s12888-019-2152-1.

78. **Byun S**, Kim AY Jang EH, Kim S, Choic KW, Yu HY, Jeon HJ. Entropy analysis of heart rate variability and its application to recognize major depressive disorder: A pilot study. *Technology and Health Care* 2019; **27**: S407–S424, DOI 10.3233/THC-199037.

79. **Byun S**, Young Kim A, Jang EH, Kim S, Choi KW, Yu HY, Jeon HJ. Detection of major depressive disorder from linear and nonlinear heart rate variability features during mental task protocol. *Computer Biol Med* 2019; **112**: 103381. https://doi.org/10.1016/j.compbiomed.2019.103381.

80. **Khandoker AH**, Luthra V, Abouallaban Y, Saha S, Ahmed KI, Mostafa R, Chowdhury N, Jelinek HF: Predicting depressed patients with suicidal ideation from ECG recordings. *Med Biol Eng Comput* 55: 793-805. 2017.

81. **Lemogne,C.**,Thomas,F.,Consoli,S.M.,Pannier,B.,Jégo,B.,andDanchin,N. Heart rate and completed suicide : evidence from the IPC cohort study. *Psychosom. Med*. 73, 731–736. 2011. doi:10.1097/PSY.0b013e3182365dc7

82. **Llamocca PP**, Junestrand M, Cukic M, Urgeles D, Lopez V. Data Source Analysis in Mood Disorder Research. In: XVIII Conference of the Spanish Association of Artificial Intelligence. 2018 Presented at: CAEPIA; October 23-27, 2018; Granada, Spain.

83. **Bozkurt A**, Barcin C, Isintas M, Ak M, Erdem M, Ozmenler KN. Changes in Heart Rate Variability before and after ECT in the Treatment of Resistant Major Depressive Disorder. *Isr J Psychiatry Relat Sci* 2013; **50**: 40-46.

84. **Jaworska N**, Wang H, Smith DM, Blier P, Knott V, Protzner AB. Pre-treatment EEG signal variability is associated with treatment success in depression. *NeuroImage: Clinical* 2018; **17**: 368-377.

85. **Schultz SK**, Anderson EA, Van De Borne P. Heart rate variability before and after treatment with electroconvulsive therapy. *J Affect Disord* 44(1): 13-20. (1997).

86. **Gandahar BN**, Subbakrishna DK, Janakiramaiah N, et al. Post-seizure EEG fractal dimension of first ECT predicts antidepressant response at two weeks. *J. Affect. Disord* 52(1-3), 235-238 (1999).

87. **Lebliecka K**., Zuchowicz, U., Wozniak-Kwasniewska, A., Szekely, D., Olejarczyk, E., and David, O.. Complexity analysis of EEG data in persons with depression subjected to transcranial magnetic stimulation. *Front Physiol*. 9:1385. (2018) doi: 10.3389/fphys.2018.01385

88. **Brunoni AR**, Kemp AH, Dantas EM, Goulart AC, Nunes MA, Boggio PS, *et al.* Heart rate variability is a trait marker of major depressive disorder: evidence from the sertraline vs electric current therapy to treat depression clinical study. *Int J Neuropsychopharmacol* 2013; **16**: 1937-1949.

89. **Antal, A**., Alekseichuk, I., Bikson, M., Brockmöller, J., Brunoni, A. R., Chen, R., et al. Low intensity transcranial electric stimulation: safety, ethical, legal regulatory and application guidelines. *Clin. Neurophysiol*. 128, 1774–1809. 2017. doi: 10.1016/j.clinph.2017.06.001